# Water on graphene: Review of recent progress

C. Melios[1, 2], C. E. Giusca[1], V. Panchal[1], and O. Kazakova[1]*

[1]National Physical Laboratory, Teddington, TW11 0LW, UK
[2]Advanced Technology Institute, University of Surrey, Guildford, GU2 7XH, UK

* Corresponding author e-mail: olga.kazakova@npl.co.uk

**Abstract**

The sensitivity of graphene to the surrounding environment is given by its π electrons, which are directly exposed to molecules in the ambient. The high sensitivity of graphene to the local environment has shown to be both advantageous but also problematic for graphene-based devices, such as transistors and sensors, where the graphene carrier concentration and mobility change due to ambient humidity variations. In this review, recent progress in understanding the effects of water on different types of graphene, grown epitaxially and quasi-free standing on SiC, by chemical vapour deposition on $SiO_2$, as well as exfoliated flakes, are presented. It is demonstrated that water withdraws electrons from graphene, but the graphene-water interaction highly depends on the thickness, layer stacking, underlying substrate and substrate-induced doping. Moreover, we highlight the importance of clear and unambiguous description of the environmental conditions (i.e. relative humidity) whenever a routine characterisation for carrier concentration and mobility is reported (often presented as a simple figure-of-merit), as these electrical characteristics are highly dependent on the adsorbed molecules and the surrounding environment.

# 1    Introduction

Since its discovery by Novoselov *et al.* [1], graphene has received a great level of attention from both the research and industry communities due to its exceptional electronic properties[2]. Specifically, owing to the outstanding resistance modulation by external electric field and high electron mobility, graphene has a great potential in high-speed transistors[3], while its high optical transparency coupled with its high conductivity paved the way for optoelectronic devices, such as touch screens[4]. Most importantly, because of its true two-dimensional nature and high surface-to-volume ratio, graphene has already been used as a sensing material for gases[5], bio[6] and magnetic fields[7]. Because of the atomically flat nature of graphene, Xu *et al.* visualised for the first time the water adlayers on mica using atomic force microscopy (AFM)[8]. Schedin *et al.* reported the first graphene-based gas sensor, with single molecule detection limit, where the adsorbed gas molecules modulated the charge carrier concentration and thus the resistance of the graphene device[9]. Since then, a large number of experimental and theoretical works studied the interaction of graphene with molecules such as $NO/NO_2$[9]–[11], CO[9], [11], $O_2$[12], $NH_3$[9] and $SO_2$[13]. Covalent functionalisation of graphene has been achieved with biomolecules such as aryl diazonium salts [14] and [15], where delocalised electrons are transferred from the graphene to the molecule. The capability for gas sensing and the sensitivity to the surrounding environment is given by graphene's π electrons, which are directly exposed to adsorbed molecules. The high sensitivity of graphene to the local environment has shown to be both advantageous but also problematic for graphene-based devices, such as transistors and sensors, where the graphene resistance will change due to humidity variations. In addition to water interacting with graphene by changing its electronic properties, graphene and graphene oxide has been proven to be a suitable material for water desalination, thus paving the way for clean water availability to remote locations[16]–[18].

In this review, recent progress on the water-graphene interaction and the effect of water on graphene's electronic properties is presented, with many of the discussed results obtained in our laboratory. The aim of this review is to elucidate the interaction of water with graphene and compare the response of different types of graphene (i.e. chemical vapour deposition (CVD) and epitaxially grown graphene) synthesised using scalable methods. We also explain the effect of substrate on graphene-water interaction by considering different substrates (i.e. $Si/SiO_2$, SiC(0001), $H_2$-passivated SiC(0001)) and outline the importance of thorough encapsulation of graphene-based devices as well as the appropriate calibration of gas/humidity sensors. Moreover, we highlight the importance of clear and unambiguous description of the environmental conditions whenever a routine characterisation for carrier concentration and mobility is reported (often presented as a simple figure-of-merit), as these electrical characteristics are highly dependent on the adsorbed molecules and the surrounding environment.

## 2      Water-graphene interactions

The electronic properties of graphene (i.e. carrier concentration, mobility, resistance and work function) were found to be susceptible to adsorbed molecules and variations of the environmental conditions[10], [19]. In particular, water and other species, found in ambient air are adsorbed by graphene[20]. However, the graphene-water interactions are highly dependent on the hydrophilicity of the graphene surface. For example, clean mono-layer graphene is considered hydrophilic, however, as the number of graphene layers increases, the graphene becomes more hydrophobic[21], [22]. Nevertheless, because of its two-dimensional nature, most of the applications require graphene to be placed on a substrate. Rafiee *et al.* showed that graphene wettability is transparent to the wetting properties of the underlying substrate[23]. However, Shih *et al.* demonstrated that graphene is only partially transparent to the wetting properties of the underlying substrate, in which case the wetting transparency of graphene breaks down when it is placed on superhydrophobic and superhydrophilic substrates[24]. The wetting behaviour of graphene will be further discussed in more details in Section 2.4. When only a few layers of graphene are concerned (i.e. 1-3LG), variations in the local carrier concentration and work function can greatly influence their wetting properties and therefore water adsorption. Since graphene thickness variations are common on large-scale graphene and water is the most abundant dipolar adsorbate under ambient conditions, it is crucial to investigate how water molecules interact with different thickness domains as well as how the charge transfer is governed by different substrates. Until now, significant effort has been dedicated to both theoretical[25]–[27] and experimental[9]–[11], [28]–[35] investigations of water on graphitic surfaces to elucidate the water-graphene interaction. Theoretical aspects will be reviewed in Section 2.2 and experimental studies in Sections 2.3, 2.5 and 2.6.

### 2.1      Graphene-based nanofluidics

The hydrophilicity of graphene and its atomic thinness ensure extremely high permeability of water, making graphene a material of great interest for nanofluidic applications, where capillaries with nanometre dimensions for molecular transport are needed. In particular, graphene has been used for the fabrication of size-selective membranes and nanopores that allow the transport of water at rates far exceeding expectations[16], [36], [37].

Water transport through graphene opens up new opportunities for multiple applications of graphene-based membranes in molecular sieving[18], [38], [39], water filtration[40], purification and desalination of water[17], [41], [42].

Both pristine graphene[43]–[47], and graphene oxide/reduced graphene oxide[47]–[51] have been used for the fabrication of graphene membranes. Control of pore size in pristine graphene, in the nm and sub-nm range is critical for the realization of high water flux and selectivity. Nanopores with finite size distribution in graphene can be induced by a variety of methods, such as electron and ion beam irradiation[52], [53] that nucleate reactive defects in graphene, which selectively grow into larger pores by chemical and plasma etching. Tuning the pore creation process demonstrates nanofiltration membranes with high permeation to water and monovalent ions[37]. Extremely fast water flow was demonstrated through artificial capillaries ingeniously fabricated using graphite crystals separated by an array of spacers made from few-layer graphene, enabling atomic precision over the channel height[54]. Water flows in channel heights, ranging from one to several dozen atomic planes, with high speed, of up to 1m/s, due to high capillary pressure of $\sim 10^3$ bar, large slip lengths, reduced friction and increased structural order in nanoconfined water.

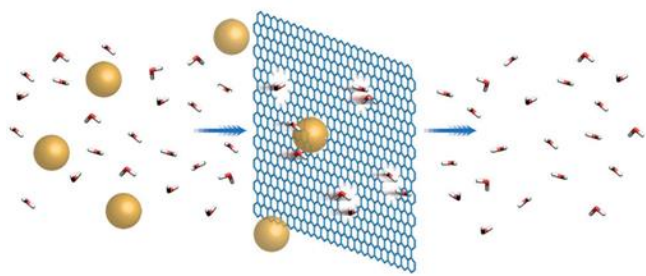

*Figure 1: Graphene membrane separates saltwater subjected to high pressure by driving water molecules through nanoporous graphene while blocking the salt ions (spheres). Reprinted by permission from Macmillan Publishers Ltd: Nature Nanotechnology* [42], *copyright (2012).*

Due to the low-friction flow of a monolayer of water through two-dimensional capillaries formed by closely spaced graphene oxide sheets, it was observed that graphene oxide membranes allow unconstrained permeation of water, while they completely block permeation of liquids, vapours, and gases, including helium[16]. Further, chip-integrated graphene oxide membranes have been shown to enhance flow control on centrifugal microfluidic platforms, allowing the passage of water, while blocking pressurised air and organic solutions[55]. Indeed, ultrahigh water permeability in nanoporous graphene has been predicted by a number of molecular dynamics (MD) studies, which investigate the physical mechanisms underlying the remarkable behaviour of water transport through graphene nanopores[36]. MD studies find that the water flux depends on the density, size and geometry of nanopores[56]. The chemistry of nanopores is also an important factor, as hydroxyl groups at the edges of graphene pores almost double the water flux due to their hydrophilic character[57]. The rapid flow is generally attributed to a lack of friction between the water and the pore, although the underlying reason is still not clear[58], therefore more research is required for a full understanding of the transport mechanism of water in nanoporous graphene to lead future developments in this area.

## 2.2    Modelling of water-graphene interactions

Density functional theory (DFT) calculations show contradictions when different graphene-water systems are considered. Some studies showed that adsorbed water has very little effect on the electronic structure of pristine graphene[59], whereas others demonstrated that different charge transfer between the $H_2O$ molecule and graphene could occur for different orientations of the water molecule[11], [60]. Even when the water molecule location and a number of water molecules were considered, the outcomes varied dramatically[25], [60]. The formation of an energy gap of the order of 20-30 meV when graphene is fully covered with water molecules was also predicted by Ribeiro *et al.* [61], demonstrating that the band structure of graphene can be significantly altered due to adsorbed molecules.

Leenaerts *et al.* demonstrated that charge transfer from graphene to water occurs when the H atoms point towards the graphene, but charge transfer is reversed when the O atom points to the graphene[11]. In a subsequent work[60], Leenaerts showed that charge transfer from a single water molecule to the graphene would have an insignificant effect on the carrier concentration, and therefore the resistance of graphene will not change significantly. However, in contrast to a single water cluster, where the water dipole moments have a small average dipole moment when considering a large concentration of water molecules (ice-like formation), the dipole moments of the individual water molecules accumulate, leading to a larger effective doping of graphene[60]. In another study by Freitas *et al.*[25], similar observations were made. In their DFT studies, it was established that the orientation of water molecules has an influence on the charge transfer mechanism, such that, for small water aggregates with configurations where the oxygen atom is pointing toward the surface, the charge transfer occurs preferentially from water to graphene[25]. However, the same study also shows that for larger adsorbed clusters, charge transfer systematically occurs from graphene to water. In a more recent work, Ho *et al.* demonstrated that the orientation of the water molecule with respect the underlying graphene also depends on the initial charge (when gated) of graphene layer[62]. When negatively charged graphene was considered, the OH bonds pointed towards the graphene layer, whereas the opposite effect was observed for positively charged graphene[62]. The orientation of the water molecule with respect to the graphene layer will, of course, influence the overall dipole moment and therefore the effective doping of graphene. The influence of water orientation on the average doping of graphene was also demonstrated recently both experimentally and theoretically by Hong *et al.* [31]

Furthermore, Wehling *et al.* studied a more realistic scenario, where the graphene is placed on a $SiO_2$ substrate[59]. In this study, Wehling *et al.* demonstrated that graphene placed on a defective substrate (i.e. $SiO_2$), is more likely to be affected by water and that the underlying substrate can influence the effects of water on graphene strongly by creating dipole moments[59] (Figure 2).

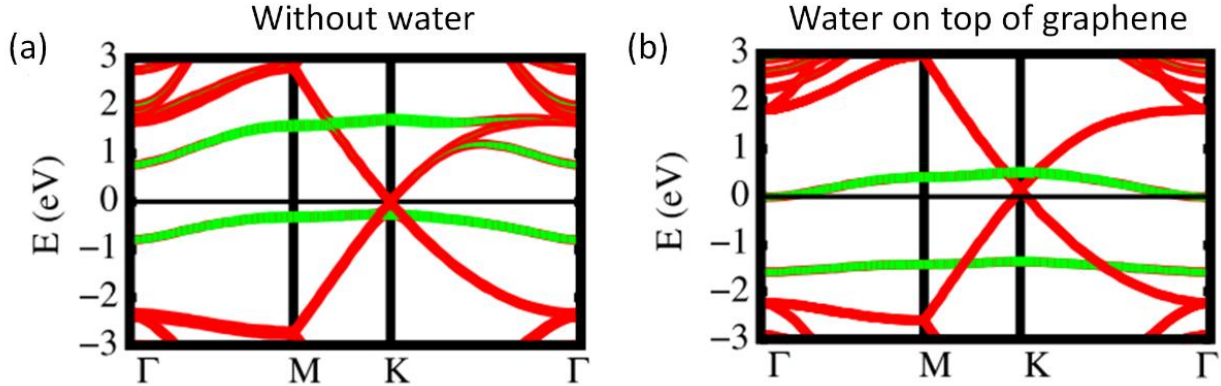

*Figure 2: Band structures of graphene on a defective $SiO_2$ substrate (a) without water and (b) with water on top of graphene, indicating moderate hole doping. Red and green lines indicate graphene and defect states contributions, respectively. Reprinted with permission from Ref. [59]*

More recent theoretical studies focused on simulating the wetting behaviour of graphene. Using Born-Oppenheim quantum molecular dynamics (QMD), Li *et al.* calculated the graphene water contact angle (WCA) of 87°[63]. Furthermore, Shih *et al.* demonstrated that the complete wetting transparency of graphene to the underlying substrate is only valid in certain cases ($30° < \theta_{substrate} < 90°$), whereas in the cases of superhydrophobic and superhydrophilic substrates, this assumption is no longer valid[24]. More complex systems were recently simulated by Driskill *et al.*[64], where the graphene WCA was found to be 7° lower, when a water layer was placed underneath the graphene, compared to water present on top. Last but not least, the work of Song *et al.* demonstrated that water confined between two graphene layers at 4.5 Å distance donates electrons to graphene, while opposite charge transfer occurs when the distance between the graphene layers is further reduced to 4 Å[65]. The last two works are of significant importance, when real life scenarios are considered, where graphene transfer is done in ambient conditions, and water layer is likely to be trapped between the graphene layers and substrate.

Despite the extensive computational theory predicting charge transfer and wettability of graphene under different simulated conditions, it was shown that even different computational models could lead to different outcomes, such as binding energies[66]. This is an important parameter to consider when simulation calculations are conducted, therefore pairing the simulating results with experimental evidence and *vice-versa* is essential. Furthermore, more complex mechanisms are involved compared to the simple interaction of graphene with a water molecule, like for example, through the water-substrate interaction[59] and graphene ripples and corrugations[67].

## 2.3 The role of substrate in graphene-water interactions

Experimentally, it was shown that the intrinsic doping level and mobility of field effect devices vary widely and their characteristics exhibit hysteretic behaviour under ambient conditions[68]–[70]. Studies concerned with the effect of ambient air exposure on the electronic properties of graphene have primarily considered two key molecular species in the air: oxygen and water. As such, unintentional hole-type doping of graphene under the influence of ambient $O_2$ molecules, well known to exhibit a rich variety of chemical interactions with aromatic hydrocarbons was demonstrated[71]–[73]. Some of these studies show that there is also an intimate relationship between the effective hole-type doping of graphene in ambient conditions and the supporting $SiO_2$ and mica substrates, also identifying the $O_2/H_2O$ redox couple as the mechanism responsible for air doping[73]–[75]. Water, for example, proved problematic for the operation of graphene-based field effect transistors (GFET). Xu *et al.* attributed the well-known hysteresis effect of GFETs transferred on Si/$SiO_2$ substrate to a layer of water trapped between the substrate and graphene, resulting in an $O_2/H_2O$ redox process, which additionally shifts the Dirac point during gating[76]. Despite current technological advances demonstrating an ability to successfully encapsulate the graphene devices and prevent their degradation in ambient humidity [77]–[79], it is still crucial to eliminate any trapped water layer between substrate and graphene (by potentially using high-temperature annealing in inert gas or vacuum)[29].

Furthermore, it was found that factors such as the type of graphene used (mechanically exfoliated, CVD or epitaxially grown, graphene oxide (GO) and reduced GO (RGO)), its thickness[22], [23], layer stacking[29] and the underlying substrate[31], [59], [80] are crucial factors for the different response of graphene to water. It was shown that the response of epitaxial graphene on SiC, as measured by changes in work function and carrier concentration, is strongly dependent on its thickness, with one-layer graphene (1LG) being the most sensitive to water adsorption and change in the environment[10], [28].

On the contrary, in the case of CVD grown graphene, the effect of thickness (i.e. non-*AB*-stacked two-layer graphene (2LG)) is not that pronounced when considering work function and carrier concentration sensitivities to water[29]. Moreover, in the case of GO and RGO it was demonstrated that their response to water is largely independent of material thickness, however, the supporting substrate plays a crucial role in the interaction with water, with Pt making both GO and RGO insensitive to humidity variations[81]. Borini *et al.* exploited the sensitivity of GO to water by developing an ultrafast GO flexible humidity sensor with 30 ms response and recovery times[30]. The fabrication process for such sensor involves simply spraying the GO flakes on a polyethylene terephthalate (PET) substrate, with resulting performance identical to a commercial humidity sensor (Figure 3a). Furthermore, Smith *et al.* demonstrated a humidity sensor developed using CVD-grown graphene transferred on Si/$SiO_2$ substrate (Figure 3b). In this work, the sensor was tested in the range of 1-95% relative humidity (R.H.) and demonstrated 0.6 and 0.4 s response and recovery times, respectively (at room temperature). The simple design of this resistive humidity sensor can offer a scalable and yet low-cost technology capable of integration with back-end-of-the-line semiconductor technologies[32]. An alternative design for GO humidity sensor was demonstrated by Bi *et al*.[82]. Their design involved a capacitive device, employing GO as the sensing material. At room temperature in the range of 23-86% R.H., 10.5 s response time was realised for this prototype, though the recovery time was relatively long (41 s)[82].

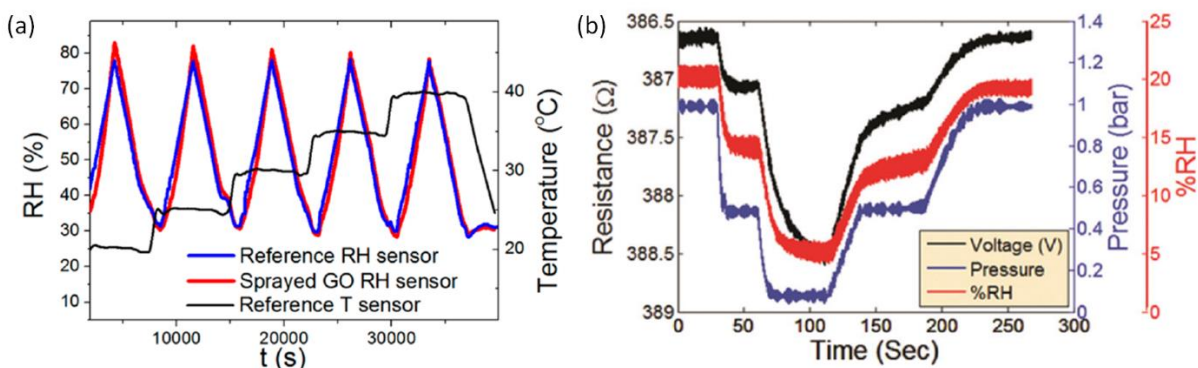

*Figure 3: (a) Relative humidity (R.H.) sensing comparison of the 15 nm thick GO sensor (red) to a commercial R.H. sensor (blue). The black line is the temperature using a reference thermometer. Reprinted with permission from [30]. Copyright (2013) American Chemical Society. (b) Resistance change as a function of time of the CVD graphene sensor (black) compared to the commercial HIH-4000 humidity sensor (red). The pressure in the chamber is also plotted as the blue line. Reprinted with permission from [32].*

## 2.4    Wettability of graphene

The wetting properties of graphene were investigated since the isolation of the first flakes, with results ranging from hydrophilic to hydrophobic graphene, without a clear, unambiguous answer. However, the reason for this discrepancy is due to several fundamental factors, such as the thickness of graphene [22], underlying substrate[23] as well as doping induced by substrate[31], [80], [83]. Optical methods and X-ray reflectivity, as well as molecular dynamics and DFT calculations, revealed a hydrophobic character for graphene as evidenced on the macroscale by a large contact angle, up to 93°, between a water droplet and graphene surface[60], [84]–[87]. On the contrary, studies involving WCA measurements coupled with infrared spectroscopy and X-ray photoelectron spectroscopy investigations showed that intrinsic graphene is mildly hydrophilic but the exposure to volatile hydrocarbons commonly present in air lowers its surface energy, making it more hydrophobic[88]. Graphene's wetting properties were also studied by Rafiee *et al*. by measuring the WCA between various substrates and graphene[23]. In this study, it was found that the contact angle between graphene-coated substrates (Si, Au, and Cu) and water only changes slightly compared to the bare substrates (Figure 4a), highlighting the wetting transparency of graphene. The WCA, however, increased when thicker graphene was overlaid on the substrates so that the coating became more hydrophobic (Figure 4b)[23]. This demonstrates that, regarding its affinity to water, a multilayer graphene surface will behave as graphite. However, as the number of layers decreases, i.e. down to 1LG and 2LG, the graphene-coated surface will become less hydrophobic. It was also postulated that the contact angle of graphene is dependent on both the liquid-graphene and liquid-substrate interaction, resulting in different degrees of wetting transparency of graphene[89]. As such, graphene is more transparent to wetting on hydrophilic substrates, but opaquer to wetting on hydrophobic substrates.

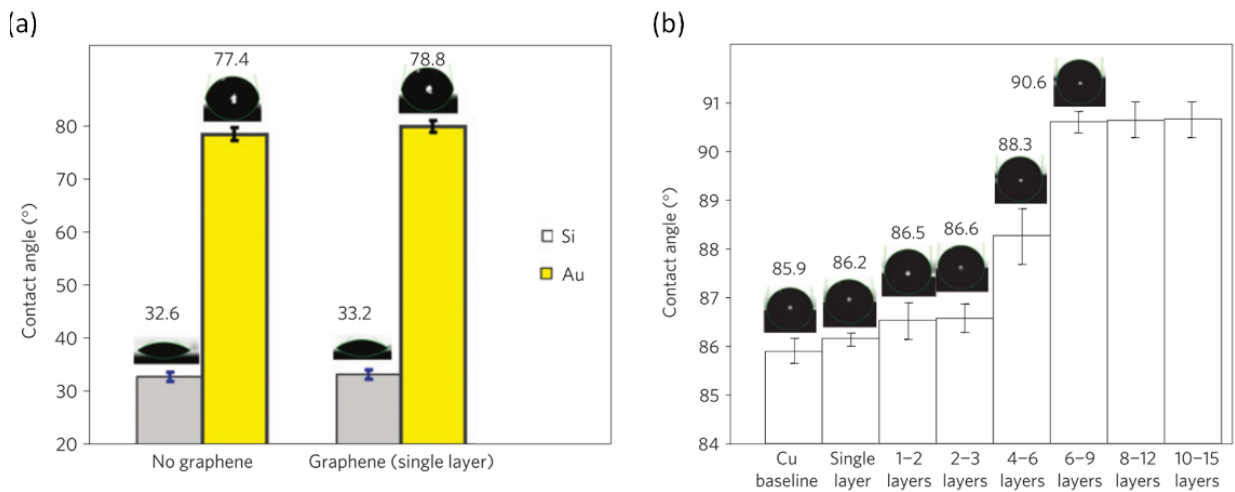

*Figure 4: (a) Water contact angle (WCA) measurements of graphene-coated (yellow) and pristine (grey) Si and Au substrates. (b) WCA measurements of Cu substrate and Cu covered with graphene of different thicknesses. Reprinted by permission from Macmillan Publishers Ltd: Nature Materials [23] copyright (2012).*

The importance of substrate and particularly of substrate-induced doping was highlighted in a recent work by Hong *et al.*, where the hydrophilicity of CVD graphene was modulated by applying an electric field (back gating)[31]. In these experiments, the authors showed that the WCA between graphene and the $Si/SiO_2$ substrate could be tuned from 78° to 60° simply by applying a back gate, and thus shifting the Fermi level from n- to p-type (Figure 5). The important implication of this is the experimental demonstration that the wettability of graphene is strongly dependent on the doping induced by the underlying substrate, even if structurally there are no changes[31]. Considering both the measurements on different substrates and identical substrate (where the doping is varied by gating) it is clear that the wetting properties of graphene greatly depend on both the wetting properties of the underlying substrate as well as the doping induced by it.

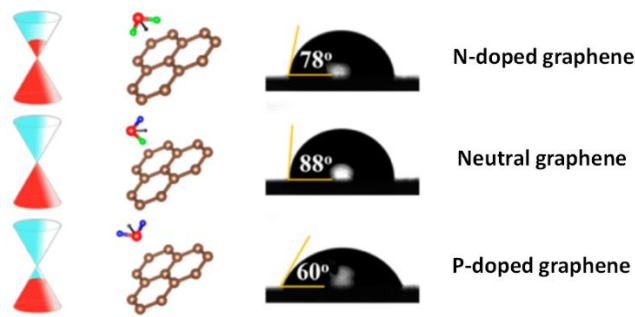

*Figure 5: Shifting the Fermi level of CVD graphene on Si/SiO$_2$ from n- to p-type leads to the change of WCA between graphene and water from 78° to 60°. Reprinted with permission from* [31]. *Copyright (2016) American Chemical Society.*

As described before, due to the simplicity of WCA measurements, they are the most commonly used technique to evaluate wettability of graphene. Nevertheless, WCA measurement can only be done on a macroscopic scale, thus averaging graphene domains of different size and not accounting for thickness inhomogeneity and defective structure. To overcome this problem, Munz *et al*. employed a technique based on chemical force microscopy (CFM) to establish the local level of hydrophobicity by measuring force-distance curves occurring between a chemically functionalised (covered with a hydrophobic layer) scanning tip and the epitaxial graphene/SiC(0001) sample immersed in de-ionised water[22]. The results presented in Figure 6 demonstrate a smaller adhesion force between the 1LG and hydrophobic tip compared to 2LG.

The differences in water adsorption for 1LG compared to 2LG are linked to changes in graphene's physical and chemical properties. The observation from adhesion mapping measurements that 1LG is less hydrophobic than 2LG is consistent with work function measurements presented in Section 2.6, showing increased water sensitivity with decreasing graphene thickness, with 1LG being the most sensitive to water. The different levels of doping-induced by the substrate, depending on the graphene thickness, give rise to surface potential and work function variations which, in turn, impact the adsorption of molecules and the wetting behaviour of 1 and 2LG. We speculate that the SiC substrate and the buffer layer at the graphene - SiC interface, controlling the electrostatic conditions of epitaxial graphene, play a similar role in mediating the interaction of water with epitaxial graphene. The difference in the electronic structure between 1LG and *AB*-stacked 2LG and the screening of the substrate interactions (in thicker graphene) are linked to the difference in local hydrophobicity[22], [28], in agreement with previous studies indicating that graphene wettability is transparent to the wetting properties of the underlying substrate[23].

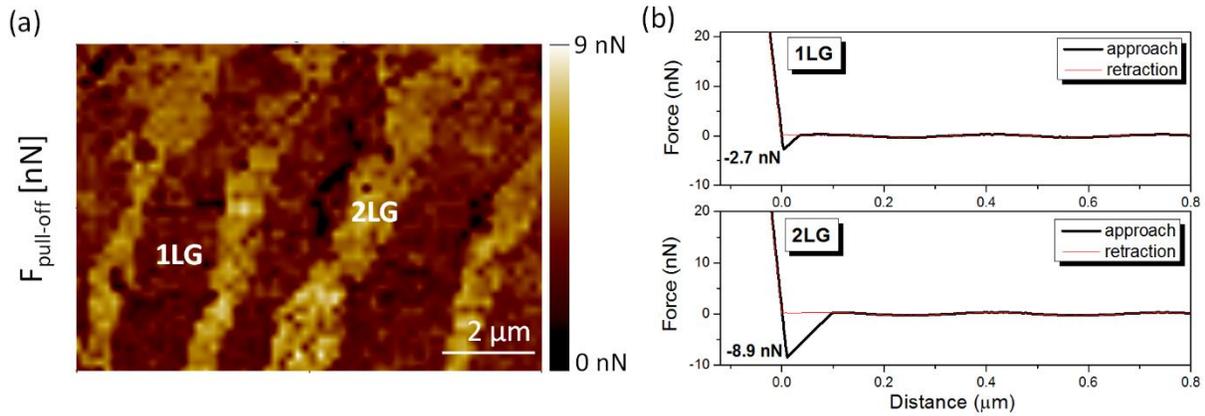

*Figure 6: (c) Adhesion map and (b) force-distance curves of epitaxial graphene on SiC(0001) obtained by chemical force microscopy*[22].

## 2.5 Water effects on global transport properties of graphene

Several groups studied the effects of water vapour on the electronic properties of graphene by measuring the changes in resistance of graphene-based devices while introducing humidity into a chamber (Figure 7). However, because the changes in resistance of graphene is a combination of carrier concentration and mobility, it only provides indirect proof of the doping induced by water (as a decrease in resistance can also originate purely from increased in charge carrier scattering). Therefore, in our previous works, we performed a comprehensive study including independent characterisation of the carrier concentration ($n$) and carrier mobility ($\mu$) by performing magneto-transport measurements in the van der Pauw geometry in highly controlled environments, starting from vacuum up to high humidity levels. Moreover, we investigated the humidity effects of different types of graphene on various substrates, namely: (i) as-grown epitaxial graphene on SiC(0001)[28], (ii) quasi-free standing graphene on SiC(0001)[83] and (iii) CVD grown graphene transferred on Si/SiO$_2$[29] (see Figure 8a-c for schematic structures). Figure 8d and Table 1 summarise the results of these studies. The epitaxial graphene on SiC exhibits an electron concentration $n_e$=3.10×10$^{12}$ cm$^{-2}$ in ambient (~23 °C, R.H. ~35%). The origin of electron doping in epitaxial graphene was attributed to the charge transfer from the interfacial layer (IFL)[90]. In contrast with epitaxial graphene on SiC, the QFS 1LG is intrinsically p-doped ($n_h$=6.43×10$^{12}$ cm$^{-2}$) under ambient conditions, due to the spontaneous polarisation of the 4$H$-SiC substrate [91], [92]. CVD graphene on Si/SiO$_2$ is also p-doped in ambient ($n_h$=1.85×10$^{13}$ cm$^{-2}$), due to charges in the underlying native oxide[29].

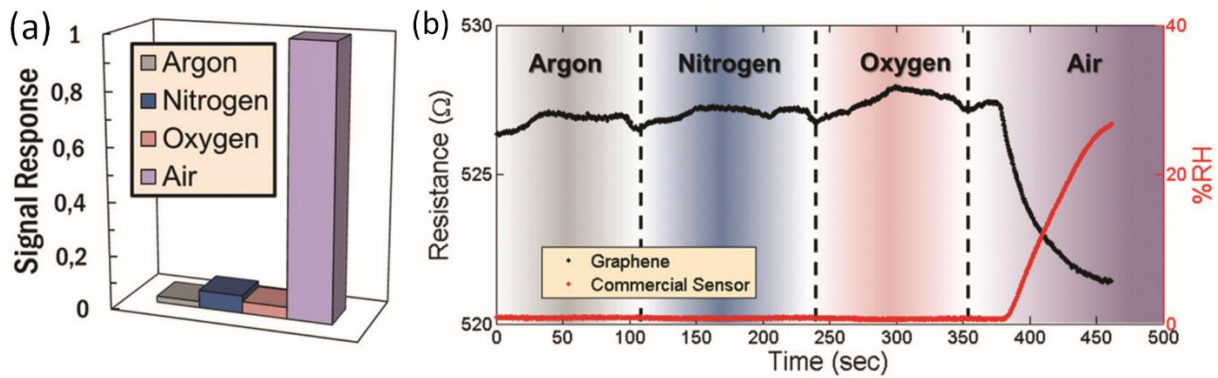

*Figure 7: Resistance changes upon exposure of a CVD graphene humidity sensor to atmospheric gases and water. Reproduced with permission from Ref. [32].*

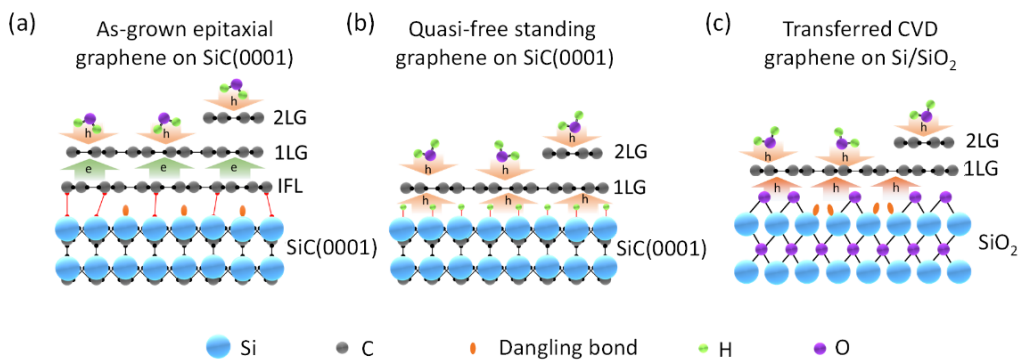

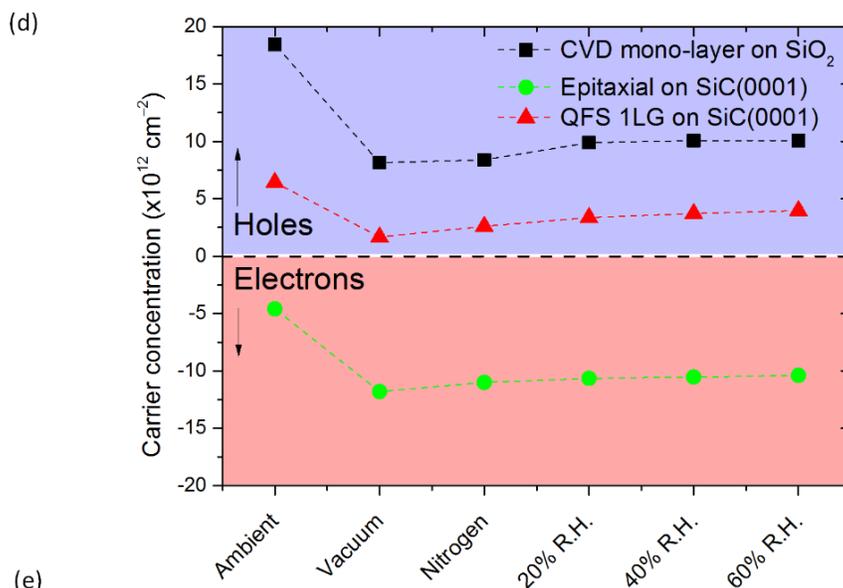

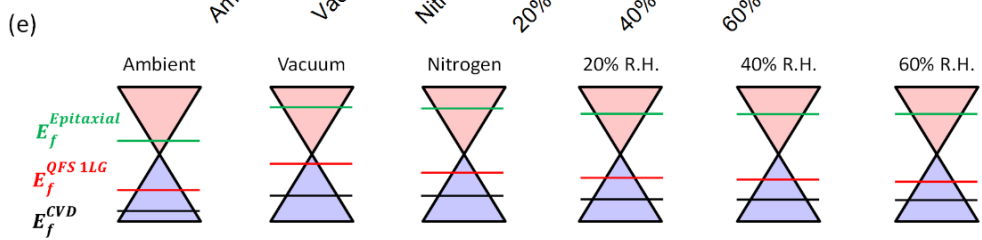

*Figure 8: Schematic representation of the graphene structure and charge transfer for (a) epitaxial, (b) QFS and (c) CVD graphene. (d) The carrier concentration of 1LG in the case of epitaxial graphene (green circles), QFS 1LG (red triangles) and CVD on Si/SiO$_2$ at various environmental conditions (ambient, vacuum N$_2$ and 20-60% R.H.). (e) Schematic representation of the Fermi energy changes for the different graphene types.*

Following vacuum annealing, airborne contaminants are desorbed from the graphene surface, increasing the electron concentration to $n_e$=1.18×10$^{13}$ cm$^{-2}$ for the case of epitaxial graphene on SiC, while for both QFS 1LG and CVD graphene the hole concentration decreases to $n_h$=1.67×10$^{12}$ cm$^{-2}$ and $n_h$=8.15×10$^{12}$ cm$^{-2}$[29], respectively. This demonstrates that in all three types of graphene, atmospheric contaminants act as p-dopants. Subsequently, nitrogen was introduced into the chamber, where a notable p-doping was observed in all cases, due to impurities transferred from the plastic pipes (despite intense flushing of the pipes).

Introducing 20% R.H. into the chamber resulted in a decrease in the electron concentration in epitaxial graphene on SiC, while the hole concentration of CVD graphene and QFS 1LG increased (Table 1). Further increase in humidity up to 60% R.H. decreased the electron concentration by a total of 0.6×10$^{12}$ cm$^{-2}$ (epitaxial graphene) and increased the hole concentration by 1.36×10$^{12}$ cm$^{-2}$ and 1.68×10$^{12}$ cm$^{-2}$ for the QFS 1LG and CVD graphene, respectively. This is a clear indication of the p-doping effect of water vapour on graphene. However, considering Table 1 and Figure 9, it is clear that the magnitude of the carrier concentration change due to water highly depends on the substrate and the relevant substrate-induced doping (as seen at the N$_2$ stage). In both cases of p-type graphene (QFS 1LG and CVD grown), a larger change in carrier concentration is observed compared to the n-type graphene on SiC. This can potentially be attributed to the intrinsic substrate-induced doping of graphene, as demonstrated by Hong *et al.*[31], where the hydrophilicity of graphene was controlled by varying the doping level using a back gate. Subsequently, changing the doping type of graphene from n- to p-type resulted in WCA between the graphene and water to 78° and 60°, respectively[31], with p-type graphene more hydrophilic than n-type graphene.

The interaction of water with graphene of different carrier type (i.e. epitaxial (*n*) and QFS 1LG (*p*)) was also investigated in our laboratory using both WCA and transport measurements, where similar conclusions were drawn (i.e. 96° for epitaxial and 86° for QFS 1LG)[83]. Nevertheless, in all three samples upon exposure to the highest humidity level (60% R.H.), the carrier concentration did not reach the values measured in ambient air (typically ~35% R.H.). This is an indication that water alone is not responsible for the doping measured in graphene in ambient environments[83]. It was demonstrated that such gases as O$_2$, NO$_2$ CO$_2$ and hydrocarbons can p-dope graphene[10]. In recent experiments, the laboratory air was measured and the presence of the most prominent contaminates, NO$_2$ at ~12.9 ppb and CO$_2$ at ~420.2 ppm, was identified. We have then synthetically reproduced these conditions in the environmental chamber and found that NO$_2$ and CO$_2$ concentrations at these levels can further decrease the electron concentration of epitaxial graphene by

~26% and 7%, respectively. However, additional traces of other gases such as NO, NOx, $O_3$ and $CH_4$, as well as organic compounds might also further dope graphene.

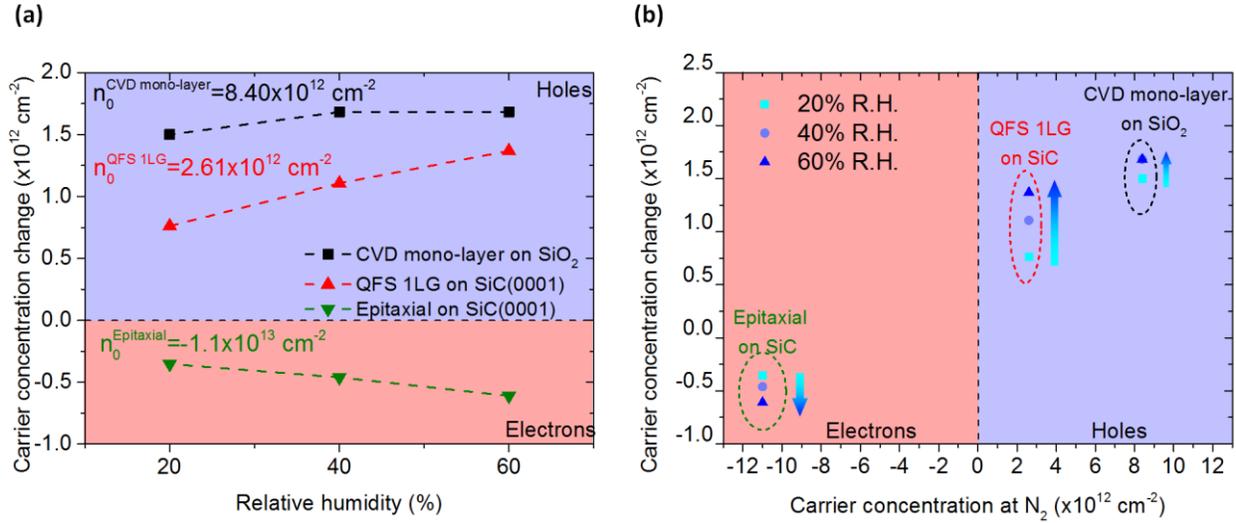

*Figure 9: (a) Carrier concentration changes for the different graphene types for 20-60% R.H. (b) Carrier concentration changes as a function of initial carrier concentration at $N_2$ for the different graphene types at various humidity levels. The arrows indicate the increase in R.H. (from light to dark blue) for each group.*

*Table 1: Summary of the humidity induced changes in the carrier concentration as compared to the pure nitrogen environment. ↑/↓ symbols indicate the increased/decreased values as compared to the nitrogen stage.*

| | Carrier concentration changes with respect to nitrogen stage ($cm^{-2}$) | | |
|---|---|---|---|
| **Relative humidity** | $\Delta n_e^{Epitaxial}$ [83] | $\Delta n_h^{QFS\ 1LG\ grown}$ [83] | $\Delta n_h^{CVD\ grown}$ [29] |
| **20%** | ↓ $0.35 \times 10^{12}$ | ↑ $0.76 \times 10^{12}$ | ↑ $1.50 \times 10^{12}$ |
| **40%** | ↓ $0.46 \times 10^{12}$ | ↑ $1.10 \times 10^{12}$ | ↑ $1.68 \times 10^{12}$ |
| **60%** | ↓ $0.61 \times 10^{12}$ | ↑ $1.36 \times 10^{12}$ | ↑ $1.68 \times 10^{12}$ |

Water not only affects graphene through changing the carrier concentration. As it was recently demonstrated by our group[83], water is also responsible for changes in the carrier mobility. This is particularly important for standardisation procedures were simply quoting the carrier mobility and carrier concentration is not sufficient; one must also specify the environmental conditions in which the measurement are conducted. The experiments reported in Figure 10 were performed on epitaxial and $H_2$-intercalated (QFS 1LG) graphene on SiC(0001) using magneto-transport measurements in identical van der Pauw devices. Figure 10a shows the carrier mobility-carrier concentration ($\mu$-$n$) relation for various humidity levels for the epitaxial graphene. To help with this investigation, Table 2 was constructed by calculating the slopes of $n$-$\mu$ plots of figure 10 for the cases of constant humidity levels and increase in humidity (red stars from figure 10). While it is expected that with increasing carrier concentration, the mobility should decrease due to increase in carrier scattering[93], [94], the opposite trend was reported, with mobility decreasing with a decrease (increase) in carrier concentration (relative humidity). This is a clear indication that in epitaxial graphene on SiC, water acts both as a Coulomb and impurity scattering centre, decreasing the mobility. This reveals two competing mechanisms that are responsible for carrier mobility changes under water-induced doping of graphene on SiC (Table 2): i) increase in mobility due to a decrease in electron concentration, ii) decrease in mobility due to the presence of water layer (schematic illustration of the different doping mechanisms is shown in Figure 8a). In the case of as-grown epitaxial graphene, the latter mechanism dominates by decreasing the mobility ~24 cm$^2$/Vs (within the range of 20-80% R.H.). However, in the case of the p-doped QFS 1LG (Figure 10b), the presence of water further increases the hole concentration and the two mechanisms of carrier scattering and Coulomb and impurity scattering (due to water) act together to decrease the mobility by ~76 cm$^2$/Vs (~3 times larger effect as compared to the epitaxial graphene).

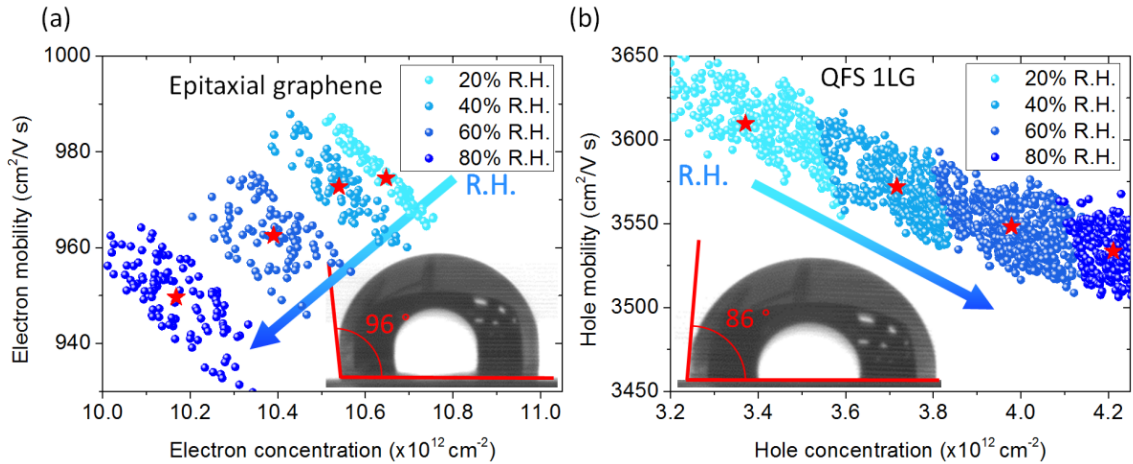

*Figure 10: Carrier mobility as a function of carrier concentration for (a) epitaxial and (b) QFS graphene on SiC for various humidity levels [83]. Red stars indicate the average value for carrier concentration and mobility at each R.H. step. The spread in the individual data points of the same colour are due to the fluctuations (with time) in the carrier concentration and therefore carrier mobility during the magneto-transport measurements. The standard deviation for carrier concentration and mobility for epitaxial 1LG (QFS 1LG) is ±0.8×10$^{11}$ cm$^{-2}$ (±1.6×10$^{11}$ cm$^{-2}$) and ±7 cm$^2$/Vs (±18 cm$^2$/Vs), respectively.*

*Table 2: Slope of the mobility – carrier concentration plots (μ-n) for the as-grown and QFS 1LG graphene on SiC(0001). The slope of μ-n at a constant R.H. was extracted by fitting the individual (all blue) data points of Figure 10. The slope of the μ-n for increase in R.H. was extracted by fitting the average points of Figure 10 (red stars).*

|  | Slope of *μ-n* plot | |
|---|---|---|
|  | **As-grown epitaxial** | **QFS 1LG** |
| **Increase in *n* at constant R.H.** | $-3.55 \times 10^{-11}$ | $-13.3 \times 10^{-11}$ |
| **Increase in R.H.** | $+5.37 \times 10^{-11}$ | $-8.98 \times 10^{-11}$ |

## 2.6   Effects of water on the local electronic properties of graphene

Having discussed the effects of water on the global transport properties of graphene, it is important to understand how these changes occur on the local scale, to clarify the role of graphene thickness and local defects. One of the earliest studies investigating the graphene-water interactions was done by Moser *et al.* using mechanically exfoliated flakes and electrostatic force microscopy (EFM)[95]. In these experiments, it was found that adsorbed water molecules on top of graphene flakes on $Si/SiO_2$ generated local dipoles, which shift the work function of graphene[95]. Verdaguer et al. [96] demonstrated another experiment using scanning probe microscopy, where the discharging of graphene flakes on $Si/SiO_2$ was found to depend on the ambient humidity levels. More recently, Ashraf *et al.*[80] performed WCA and work function measurements on graphene placed on different substrates, providing either n- or p-doping. Among their results, they demonstrated that the WCA is highly dependent on the work function of the graphene layer[80].

Bollmann *et al.* studied the doping effects from layers of water trapped between exfoliated graphene and insulating $CaF_2(111)$ substrate (Fig. 11a-c)[97], [98]. Their study revealed that graphene is hole-doped from the presence of water, which draws electrons from the graphene and towards the water. A single layer of water has the largest p-doping effect, drawing as much as 0.013 holes per unit cell of graphene, whereas a second layer of water draws an additional 0.004 holes per unit cell (Fig. 11d). Beyond the thickness of approximate five layers of water, the hole doping effect is virtually saturated.

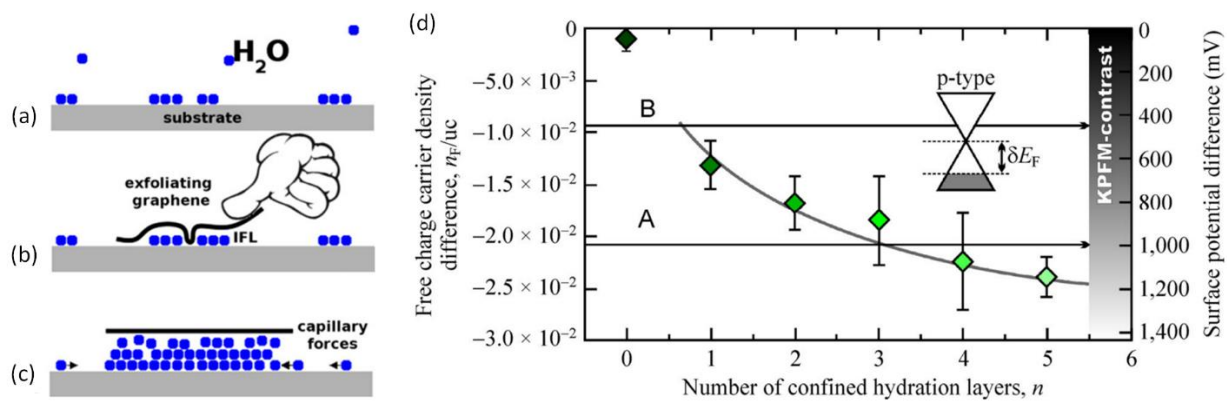

*Figure 11: Illustration of (a) water from the air adsorbing onto CaF2(111) substrate, (b) a film of water trapped between the mechanically exfoliated graphene and the substrate, and (c) capillary forces confining the film of water between graphene and the substrate. Reproduced with permission from* [97]. *(d) The relation of free charge carrier density and surface potential difference of graphene with increasing number of confined water layers. Reproduced with permission of Springer from* [98].

Shim *et al.* reported a similar observation of exfoliated graphene on mica, where the presence of water between graphene and mica lead to a significant shift of the Raman G and 2D peaks resulting in a shift in the Fermi energy of -0.35 eV (Fig. 12), which translated to hole density of $9\times10^{12}$ cm$^{-2}$[99]. The diffusion of water can also occur for graphene on $SiO_2$ substrates under high humidity environments[100], forming stable ice-like water layers which are also stable in ambient air. Observations by Jung Lee *et al.* show that highly mobile liquid phase water can also diffuse between the graphene and $SiO_2$ substrate, which is attributed to the high roughness and lower hydrophilicity of $SiO_2$ compared to mica. However, the liquid water typically causes wrinkles and can even lead to folding of graphene. This trapped water can also influence the doping effect in graphene, as reported for graphene on $CaF_2$(111) and mica substrates.

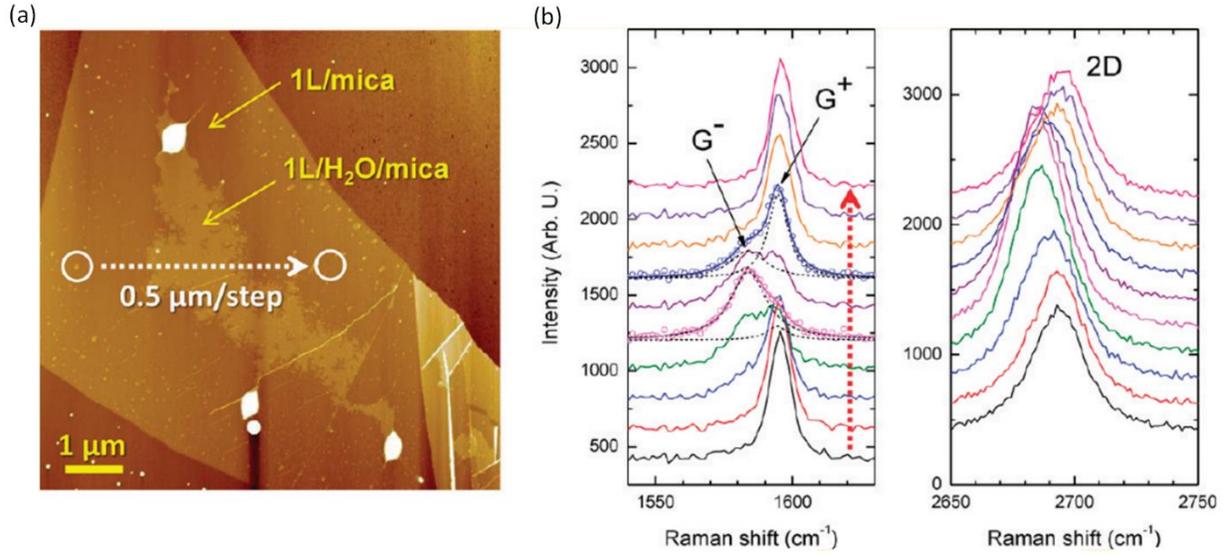

*Figure 12: (a) Topography image of graphene on mica. (b)(Right) G and 2D Raman peaks taken from graphene on mica along the arrow indicated in the left panel. Reprinted with permission from* [99]. *Copyright (2012) American Chemical Society.*

Following the work of Moser *et al.*, we advanced the measurement of local electronic properties of graphene under highly controlled environments (starting from vacuum, up to 60% R.H.), to scalable graphene types (i.e. epitaxial on SiC and CVD grown) and most importantly calibrated work function mapping. The work function maps for three samples (as-grown and QFS 1LG epitaxial graphene and CVD graphene) are presented in figure 14 and the summary of the work function difference between 2LG and 1LG is presented in Figure 15. In ambient conditions (Figure 14a-c), all three samples exhibit higher work function for 1LG compared to 2LG: $\Phi_{1LG} > \Phi_{2LG}$. Although the work function for 1LG ($\Phi_{1LG} = -E_F + eU_{CPD}$) is directly related to carrier concentration through the Fermi energy ($E_F^{1LG} = v_F \hbar \sqrt{\pi n}$, where $v_F$ is the Fermi velocity), in the case of *AB*-stacked 2LG (such as, as-grown and QFS graphene on SiC), the situation is further complicated due to more complex relation between the Fermi energy and carrier concentration. However, for CVD graphene on Si/SiO$_2$ the two layers are only weakly coupled and can be approximated as two independent layers [101]. The weak coupling of twisted 2LG islands grown by CVD is demonstrated in figure 13a, where the Raman spectra of epitaxial and CVD graphene are presented. The 1LG on SiC(0001) and Si/SiO$_2$ are very similar, featuring a symmetrical 2D-peak (figure 10a). However, the 2LG on SiC(0001) exhibits a characteristic 2D-peak of *AB*-stacked graphene, whereas 2LG on Si/SiO$_2$ resembles the one of 1LG. This is due to the weakly coupled twisted layers, which behave as electrically decoupled (with the E-k dispersion approximated as linear). Furthermore, work function measurements using FM-KPFM demonstrated that the as-grown 2LG islands in figure 13c exhibit the same work function as the artificially transferred 2LG stack of figure 13e. This indicates that the CVD grown 2LG islands are weakly coupled. The bottom (first) layer of graphene screens the substrate charges, therefore the 2LG islands are less affected by the substrate doping, therefore it exhibits a lower work function and hole concentration (compared to the underlying graphene layer)[29]. Havener et al.[102] demonstrated that it is possible to estimate the relative twist angle of the two graphene layers by measuring the normalized (to 1LG) area of G-peak. They also verified that the G-peak area shows massive enhancement for angles between ~11°-14°. However, for smaller or larger angles, the G-peak area is approximately the same, making it difficult to quantitatively establish the rotational angle between layers.

Table 3: Comparison between CVD graphene transferred on Si/SiO$_2$, epitaxial and QFS graphene on SiC(0001).

| Property | CVD graphene transferred on Si/SiO$_2$ | As-grown epitaxial graphene on SiC(0001) | QFS 1LG on SiC(0001) |
|---|---|---|---|
| Carrier type | p-type[29], [103] | n-type[83], [90] | p-type[83], [91], [92] |
| Total carrier concentration change 0-60% R.H. | ↑ $1.68 \times 10^{12} cm^{-2}$ | ↓ $0.61 \times 10^{12} cm^{-2}$ | ↑ $1.36 \times 10^{12} cm^{-2}$ |
| 2LG stacking | Weakly coupled[101], [102] | *AB*-stacked[104] | *AB*-stacked[104] |
| Hydrophilicity of 1LG and 2LG | Similar for 1LG and 2LG (_no_ swap of contrast)[29] | 1LG: more hydrophilic than 2LG (swap of contrast)[10], [22], [28], [83] | 1LG: more hydrophilic than 2LG (swap of contrast)[83] |
| Carrier concentration sensitivity to water | Similar for 1LG and 2LG[29] | 1LG more sensitive[10], [22], [28], [83] | 1LG more sensitive[83] |
| Ambient air and 80% R.H. | Water alone does not restore *Φ*, $n_h$ and $n_e$[29], [10], [22], [28], [83] | | |

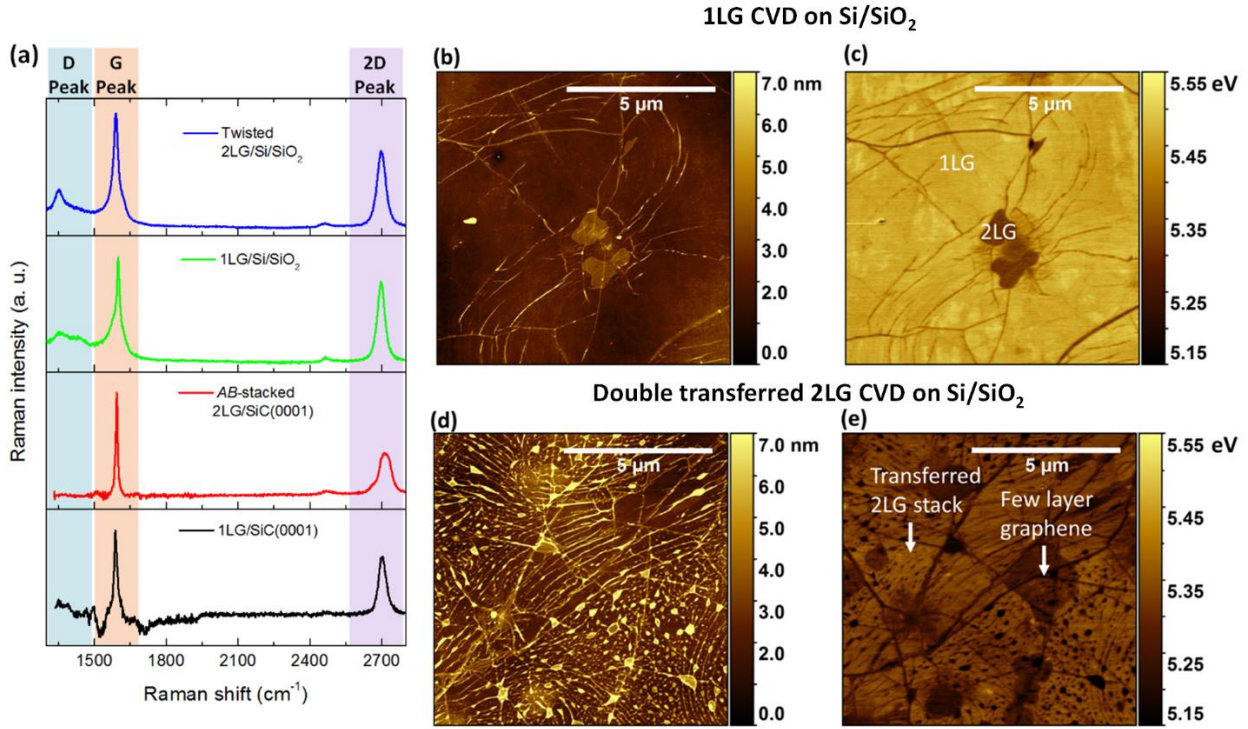

*Figure 13: (a) Raman spectra of graphene grown epitaxial on SiC(0001) (with the SiC spectrum subtracted) and using CVD, transferred on Si/SiO$_2$. The 2D-peak of 2LG is the characteristic of AB-stacked layers, however the symmetric 2D-peak of CVD 2LG islands indicates the weak coupling between the layers. (b) Topography and (c) work function maps of the 1LG, showing 2LG islands with lower work function. (d) Topography and (e) work function map of the transferred 2LG graphene stack showing contamination due to the transfer procedure and air exposure as well as thicker few layer graphene islands of lower work function (identical to as-grown 2LG in (c)).*

When the environmental chamber is evacuated to vacuum of ~ 1×10$^{-5}$ mbar in both epitaxial and QFS 1LG, the work function contrast between $\Phi_{1LG}$ and $\Phi_{2LG}$ is inverted, indicating $\Phi_{1LG} < \Phi_{2LG}$ (Figure 14d and e). However, in the case of CVD graphene, the work function of 1LG is higher than that of 2LG (Figure 14f) and the difference increases to $\Delta\Phi_{2-1LG} = 224\ meV$ (Figure 15). The increase in $\Delta\Phi_{2-1LG}$ is a clear indication that the doping between the two layers is now significantly different compared to ambient. This demonstrates that the bottom layer screens most of the substrate-induced doping (as most of the atmospheric contaminants were desorbed in vacuum) with the top layer being less affected by the substrate. Contrary to CVD graphene, in the case of 1LG and *AB*-stacked 2LG on SiC, the reversal in work function between the two layers is due to the difference of their hydrophobicity and their electronic band structure [22], [28].

Once the chamber was filled with dry N$_2$ up to atmospheric pressure, 20% R.H. was introduced into the chamber (Figure 14j-l). In this controlled environment, the work function in all three samples increased indicating p-doping, thus confirming the results of transport measurements described in section 2.5. Further increase in R.H. up to 60%, resulted in an additional increase in work function in all three samples. However, significant changes appear in the work function difference between 1LG and 2LG (Figure 15). For graphene on SiC(0001), the work function difference between the two layers approaches zero with increasing humidity, with the epitaxial graphene sample exhibiting contrast reversal again at ~60% R.H. However, for the CVD graphene on Si/SiO$_2$, $\Delta\Phi_{2-1LG}$ remains mostly constant (changed by only ~18 meV) in the range of 20-60% R.H., indicating that both 1LG and 2LG are affected similarly in this range. This is a fundamental difference between *AB*-stacked 2LG on SiC(0001) (in which case the strongly electrically coupled layers behave as single system) and weakly coupled 2LG on Si/SiO$_2$, in which case they behave as individual graphene layers. To verify this, a controlled experiment was carried out, where transport measurements were performed on an artificially constructed randomly stacked CVD grown 2LG structure, by individually transferring two graphene layers on top of each other. In this experiment, the hole concentration changed by the same amount for both 1LG and 2LG [29]. Moreover, in the case of as-grown epitaxial *AB*-stacked 2LG on SiC(0001), symmetry breaking due to difference in charge density between the top (largely exposed to p-type doping from environment) and bottom (largely exposed to n-type doping from substrate) layers can induce a small bandgap, however this is beyond the scope of this study[105].

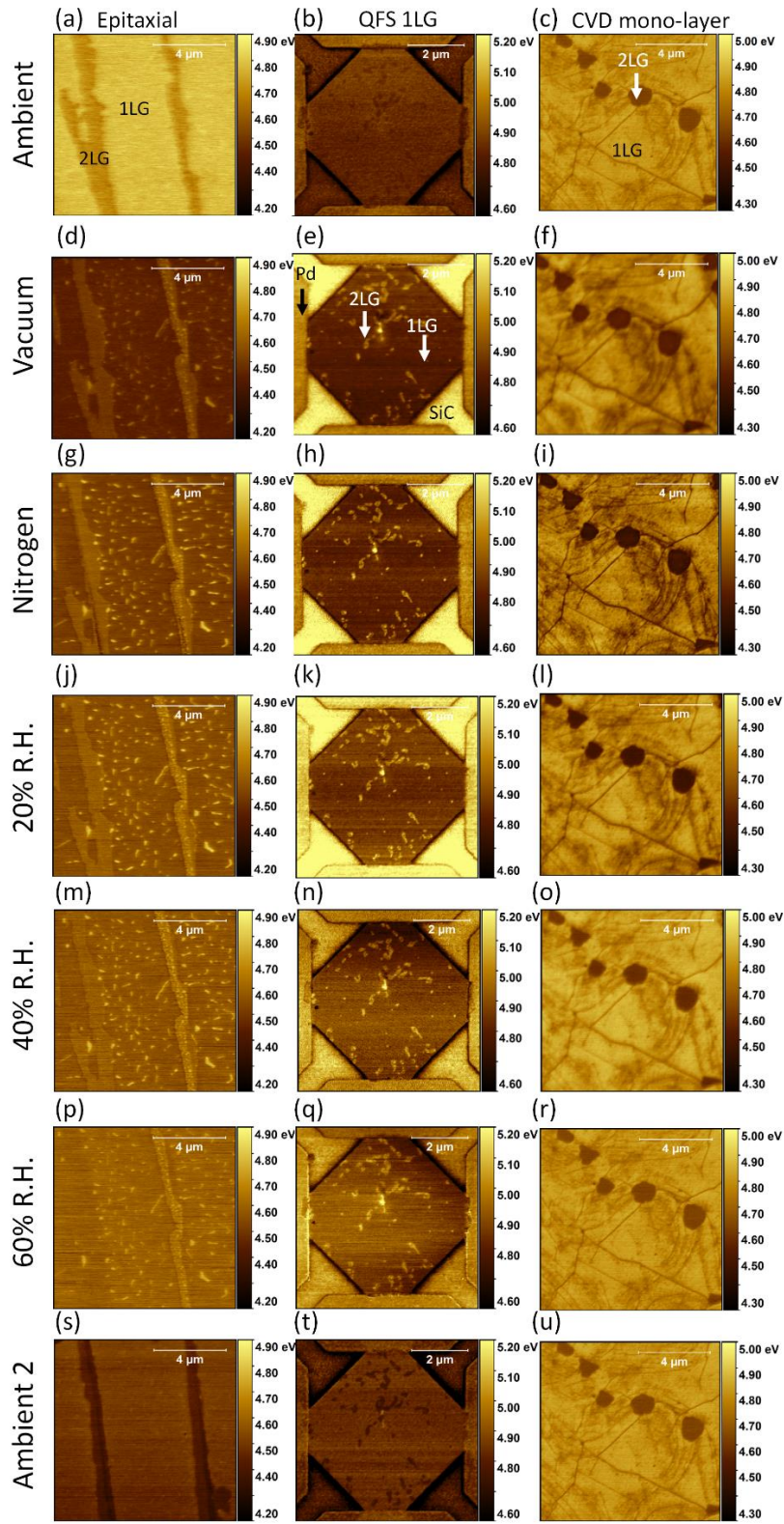

*Figure 14: Work function maps of epitaxial graphene on SiC(0001) (left column)[28], QFS 1LG on SiC(0001) (middle column)[83] and CVD graphene transferred on Si/SiO$_2$ (right column)[29]*

*for ambient (a-c), vacuum (d-f), nitrogen (g-i), 20-60% R.H. (j-r) and second ambient (s-u) environments.*

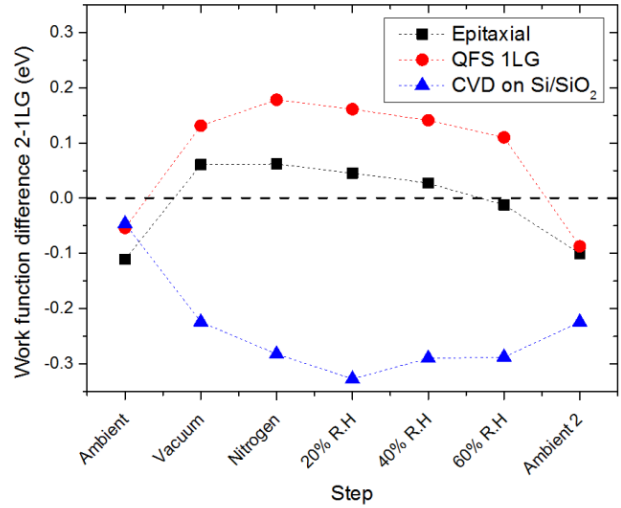

*Figure 15: Work function difference between 2-1LG for epitaxial graphene on SiC(0001) (black)[28], QFS 1LG on SiC(0001) (red)[83] and CVD graphene transferred on Si/SiO$_2$ (blue)[29] for different environmental conditions. The change of sign for the work function difference in the case of epitaxial and QFS 1LG indicates different hydrophilicity of 1LG and 2LG.*

## 3 Conclusion

Recent progress in understanding the interaction of water with graphene was reviewed, which presents recent theoretical and experimental advances. In particular, modelling of water-graphene interaction, wettability of graphene and the substrate-water effects on the local and global electronic properties of graphene were discussed.

The effects of humidity on the electronic properties of a variety of graphene types on different substrates were presented. The study of different graphenes demonstrated that substrate plays a crucial role in both the wetting and sensitivity of graphene to water vapour. Most importantly, it is suggested that the intrinsic doping of graphene, induced by the substrate, is only partly responsible for the changes in the electronic properties (i.e. carrier concentration and mobility) of graphene when water molecules are introduced on the surface. By comparing the changes in carrier concentration of epitaxial graphene on SiC(0001), QFS 1LG and CVD-grown graphene transferred on Si/SiO$_2$, we establish that both the p-doped QFS and CVD-grown graphenes exhibit higher sensitivity to water compared to n-type epitaxial on SiC(0001). Furthermore, the carrier mobility of QFS 1LG showed a greater decrease with increasing humidity. This was attributed to the combination of both charge carrier and impurity scattering, a result of both the increase in carrier concentration and the formation of a water layer, respectively. The thickness dependence of water sensitivity was also reviewed for the three graphene types. While in CVD graphene on Si/SiO$_2$, the two graphene layers are weakly coupled, resulting in very similar response to water, the work function in *AB*-stacked 2LG on SiC reveals a thickness dependence. This was attributed to the fundamental difference in the electronic structure of *AB*-stacked graphene, which exhibits a parabolic dispersion, compared to randomly stacked 2LG.

Understanding the effects of humidity/water on graphene is particularly important for graphene-based electronics, operating under different environmental conditions (*i.e.* from vacuum to low and high humidity). The results reviewed here point to a range of applications for QFS 1LG, such as humidity sensors, and the need for proper encapsulation of graphene-based devices for stable electrical behaviour. We further emphasise the importance of following a standardised procedure for the accurate characterisation of magneto-transport properties, instead of just resistance measurements, due to the combined contribution of carrier concentration and mobility in the resistance value. The environmental condition and carrier concentration value should be accompanied by the corresponding mobility value, as environmental conditions can greatly influence the electronic properties of graphene.

## 4    Methods

### 4.1    Sample preparation

#### 4.1.1.  CVD graphene on Si/SiO$_2$

Arguably, one of the most scalable solutions for large-area graphene production is growth via chemical vapour deposition (CVD) on Cu foils [106], [107]. Typically in this method, a hydrocarbon gas is flown over the Cu substrate, which is heated to ~1000 °C [107]. Here the Cu acts as the catalyst and graphene growth is initiated at the defects and grain boundaries of Cu and eventually covering the entire substrate. The graphene must be removed from Cu, which is typically done by spin coating a support layer of polymethyl-methacrylate (PMMA)and then wet etching of the Cu[29]. The PMMA layer attached to the graphene is then transferred onto a desirable substrate (i.e. Si/SiO$_2$). Other methods for graphene transfer involve dry methods [106] or electrochemical delamination [108]. Despite charge carriers in transferred CVD graphene achieving mobilities of ~2000 cm$^2$/Vs at room temperature and benefit from back-gating when the substrate is Si/SiO$_2$, the transferred graphene membrane can suffer from cracks, wrinkles, polymer residues and other transfer-induced defects, all of which can degrade the quality significantly over a large scale area[109].

### 4.1.2. As-grown epitaxial graphene on SiC(0001)

Another promising route for the preparation of wafer-scale graphene is epitaxial growth on SiC(0001) substrates. The honeycomb lattice graphene is formed on the surface of SiC after the Si atoms sublimate at high temperatures[110]–[112]. When SiC substrate is exposed to high temperatures, the surface undergoes significant changes. To begin with, the SiC(0001) is composed of a Si- and C-rich layered structure, which will then rearrange to a C-rich structure. This layer is similar to graphene, as it is composed of carbon atoms in a honeycomb lattice, but a significant amount of sp$^3$ bonds to the SiC substrate are present. This layer is called interfacial layer or buffer layer[110]. The first graphene layer will grow on top of the IFL.

Research groups are intensely trying to perfect the mechanisms of epitaxial graphene grown on SiC. When the growth is simply done under high temperature in ultrahigh vacuum (UHV), the result is an uncontrollable growth of non-uniform graphene. By introducing Ar during the annealing, the sublimation rate is suppressed. This results in better uniformity of graphene[113], [114].

Another important parameter in the epitaxial growth of graphene on SiC is the substrate preparation. Due to the miscut angle of the SiC wafer, the resulted substrate is governed by terraces and terrace edges. These edges act as nucleation sites for graphene and often lead to the formation of two-layer graphene (2LG) along the terrace edges[115].

Another growth mechanism was also proposed by Strupinski *et al*. by annealing the SiC substrates in a hydrocarbon-rich CVD reactor and flowing an Ar laminar flow to control the growth[116].

### 4.1.3. Quasi-free standing graphene on SiC(0001)

Although epitaxial graphene on SiC is generally of high mobility, it is limited by the underlying interfacial layer (IFL) and can be improved by decoupling. A favoured route to achieve decoupling of the graphene from the IFL is using hydrogen intercalation, thus forming quasi-free standing graphene (QFSG) [117]–[120]. While the principle of intercalation is very simple, in practice, achieving the optimum result is non-trivial, as parts of the sample may be partially decoupled or even etched. Annealing the graphene sample in a hydrogen rich environment at high temperatures (700–1100 °C) will enable the hydrogen to penetrate through the IFL and break the Si-C bonds, thus saturating the Si substrate and forming Si-H bonds. This decouples the non-conducting IFL from the SiC substrate and converts it into a conducting QFSG layer[117]–[120]. The results of this transformation are the change of the intrinsic doping from electrons to holes, due to the spontaneous polarisation of the SiC substrate [91], [92], and the significant enhancement of the carrier mobility [104], [117], [119].

## 4.2 Measurement techniques

### 4.2.1. Kelvin probe force microscopy

Kelvin probe force microscopy is an electrical single pass semi-contact AFM mode, which we operate by utilising frequency-modulation (FM-KPFM). In this mode, a conductive tip is oscillating at the cantilever's (Bruker PFQNE-AL with k=1.5 N/m) mechanical resonant frequency ($f_0$=300 kHz), while a much lower frequency ($f_{mod}$) of modulating AC voltage is applied to induce a frequency shift of $f_0 \pm f_{mod}$, and subsequent higher harmonics. The side lobes (monitored by a PID feedback loop) generated by this shift are minimised by applying a DC compensation voltage. By measuring the DC voltage at each pixel, a surface potential map [contact potential difference (($U_{CPD}$)] is constructed. As FM-KPFM is a force gradient technique, a high spatial resolution of <20 nm can be achieved, which is limited only by the tip apex diameter[121], [122]. This allows nanometre resolution imaging of the surface potential of graphene and provides direct information of the work function variations and number of graphene layers[104], [123]–[125]. For the calculation of the tip work function at each environmental stage, the gold contacts on the devices, or freshly cleaved Highly Order Pyrolytic Graphite (HOPG) can be used as a reference, using: $\Phi_{Tip} = \Phi_{Au} + eU_{CPD}$[29], [126]. The environmental measurements were done using KPFM equipped with an environmental chamber. For our experiments, an NT-MDT NTEGRA AURA SPM system was developed capable of performing AFM and KPFM is specific environmental conditions (vacuum, gases and humidity). The samples were initially measured in ambient conditions, following by vacuum annealing, dry $N_2$, 20-60% relative humidity (R.H.), and finally in ambient again. In each environmental step, the AFM tip was calibrated to extract the work function of each layer.

### 4.2.2. Transport measurements

For measurements in the van der Pauw geometry, the sample is placed in an electromagnet creating a magnetic field of $B_{AC}$=30 mT or a permanent magnet of $B_{DC}$=400 mT. The sheet resistance ($R_s$) is measured and calculated using $e^{\frac{-\pi R_A}{R_s}} + e^{\frac{-\pi R_B}{R_s}} = 1$, where $R_A$ and $R_B$ are the resistances obtained by passing a bias current $I_B$=50 µA and measuring the voltage drop across the opposite sides of the sample. To obtain the carrier concentration ($n = \frac{1}{eR_H}$), the Hall coefficient ($R_H$) is calculated using $R_H = \frac{V_H}{BI}$, by passing current and measuring the diagonal Hall voltage ($V_H$). The mobility is also calculated using $\mu = \frac{R_H}{R_s}$ [127]. The environmental measurements were done in an environmental chamber, by performing measurements in the following order: ambient (~23 °C, R.H. ~35%), followed by vacuum annealing, dry N$_2$, 20-60% R.H., and finally in ambient again.

### 4.2.3. Water contact angle measurements (WCA)

Water contact angle measurements can be performed by depositing water micro droplets on the graphene samples. A high magnification CMOS camera is used to measure the angle between the water and the substrate (graphene). Similar WCA measurements can be performed in an environmental scanning electron microscopy by allowing water to condensate on the sample and create bubble structures.

## 5 Acknowledgements

The authors acknowledge the support of EC grants Graphene Flagship (No. CNECT-ICT-604391), EMRP under project GraphOhm (No. 117359), EMPIR 2016NRM01 GRACE and NMS under the IRD Graphene Project (No. 119948). The work was carried out as part of an Engineering Doctorate program in Micro- and NanoMaterials and Technologies, financially supported by the EPSRC under the grant EP/G037388, the University of Surrey and the National Physical Laboratory. The authors would like to also acknowledge Wlodek Strupiński, Amaia Zurutuza, Alba Centeno, D. Kurt Gaskill and Rositsa Yakimova for providing the graphene samples used in this research and Alexander Tzalenchuk and Ivan Rungger for the useful discussions.